\documentclass[aps,pra,nofootinbib,showpacs,twocolumn,superscriptaddress,10pt]{revtex4-1}

\usepackage{natbib}
\usepackage{amsmath,amssymb}
\usepackage{graphics}
\usepackage{color}

\newcommand{\im}[0]{{\rm i}}

\newcommand{\ket}[1]{|{#1}\rangle}
\newcommand{\bra}[1]{\langle{#1}|}
\newcommand{\braket}[2]{\langle{#1}|{#2}\rangle}

\begin{document}

\title{Quantum state of a frequency comb}

\author{Filippus S. \surname{Roux}}
\email{froux@nmisa.org}
\email{stef.roux@wits.ac.za}
\affiliation{National Metrology Institute of South Africa, Meiring Naud{\'e} Road, Brummeria, Pretoria, South Africa}
\affiliation{School of Physics, University of the Witwatersrand, Johannesburg 2000, South Africa}

\begin{abstract}
Uncertainties in the frequency parameters of a frequency comb would cause it to represent a mixed quantum state. The fact that such a quantum state contains temporal frequency as an additional continuous degree of freedom, in addition to the particle-number degrees of freedom, presents challenges for its formulation. Here, we develop a formalism in terms of which one can express a mixed quantum state that contains another continuous degree of freedom, in addition to the particle-number degrees of freedom. In the expression of such a mixed state, the frequency degree of freedom is represented in terms of a power spectral density. For this purpose, we compute the power spectral density of the frequency comb laser, by taking the uncertainties in the frequency parameters into account.
\end{abstract}

\maketitle

\section{Introduction}

The development of a method to stabilize mode-lock lasers \cite{hansch} has led to significant advances in time and frequency metrology \cite{diddamsrev}. The resulting frequency comb obtained from such a stabilized mode-locked system has subsequently found many applications \cite{yerev,diddams}. These applications include some that are related to quantum information \cite{qinf0,qinf1,qinf2,qinf3}. A consequence of this connection is that, in a reciprocal fashion, quantum information technology is being applied to synchronization in time and frequency metrology \cite{giovan,quan}. An understanding of the quantum properties of frequency combs, based on an accurate description of its quantum state, can benefit time and frequency metrology through the improvement of accuracy. It can also benefit quantum information applications that are based on frequency combs.

Our focus here, is to consider the effects of uncertainties in the frequency parameters of a frequency comb on its quantum state. While its quantum state is often considered as a pure state (see for example \cite{yoon}), the uncertainties in the parameters of a frequency comb imply that the effective quantum state is a mixed state. These uncertainties are associated with the temporal degree of freedom of the state in the Fourier domain. Both the particle-number degrees of freedom, which govern the photon statistics of the frequency comb source, and the temporal degree of freedom are necessary to provide an accurate representation of the quantum state of such a frequency comb.

The popular way to represent quantum states that consist of multiple photons is to use so-called continuous variables \cite{contvar1,contvar2}, leading to the Wigner \cite{wigner}, Husimi \cite{husimi} and Glauber-Sudarshan \cite{sudarshan,glauber} representations, also referred to as quasi-probability distributions. As such, the phrase `continuous variables' refers to the complex plane over which the coherent states are defined. However, these representations only address the particle-number degrees of freedom. When the quantum state also involves other degrees of freedom, the representation of the quantum state becomes more involved.

If the additional degrees of freedom are represented by discrete values, one can use a generalization of the coherent states, composed of tensor products of coherent state elements for all the respective discrete values of the other degrees of freedom \cite{mehta}. The discrete values serve as index for the different elements in the tensor product. Such coherent states are well-defined normalized states, thanks to the discreteness of the extra degrees of freedom. The quasi-probability distributions can then be generalized in terms of such coherent states, which are defined over a tensor product space, composed a tensor product of complex planes, one for each value of the extra degrees of freedom.

This approach to generalize coherent states fails when the additional degrees of freedom are continuous, which is the case for temporal frequency, as we'll consider here. In such a context, the phrase `continuous variables' becomes confusing, because it is not only the particle-number degrees of freedom that are represented by continuous variables, but also the additional degrees of freedom.

The reason for the failure is two-fold. In the first place, one cannot form a tensor product of elements indexed by a continuous variable. The notion of such a tensor product can only be conceptualized in a rather abstract sense. Secondly, and perhaps more importantly, the individual elements of such a tensor product would not be normalizable. This follows from the fact that such an element explicitly depends on the additional degree of freedom --- all the photons in such an element carry the same value of the additional continuous variable. The Fock states that are present in the expansion of these coherent state elements would also explicitly depend on the additional degree of freedom. Their orthogonality condition would therefore need to contain Dirac delta functions for this additional degree of freedom. As a result, these Fock states cannot be normalized. In fact, the details show that they are severely divergent. The same implies to the coherent state elements composed of such Fock states.

There is an alternative way to define coherent states with additional continuous degrees of freedom that does allows them to be properly normalized. In this case, the photons in the state would all be parametrized by the same function (or spectrum, in the current context) of the additional continuous variable. However, the space over which these coherent states are defined becomes vast: instead of the two-dimensional complex plane, it is the space of all square-integrable complex functions of the additional continuous variable. An attempt to generalize the quasi-probability distributions by defining them over such a vast space would be a purely formal exercise, resulting in a formalism that would be of little use.

Here, we intend to investigate the effect of uncertainties of the frequency parameters. As a result, we need to treat the frequency as a continuous degree of freedom, in addition to the particle-number degrees of freedom. The uncertainties turn these states into mixed states. From the above reasoning, it follows that the quasi-probability distributions are not suitable for such an investigation. Hence, we first need to develop a quantum formalism in terms of which we can express the general mixed multi-photon quantum state of a frequency comb.

We'll assume that, in the absence of any uncertainties in the frequency parameters of the frequency comb, the quantum state of the laser can be expressed as a generalized coherent state, defined in such a way that it can be properly normalized. We do not consider any manipulation of the particle-number degrees of freedom, such as squeezing. The generalization incorporates the temporal degree of freedom (frequency dependence) into the definitions of the Fock states, in terms of which the coherent states can then be defined. We derive the expression for the mixed state, resulting from the effect of the uncertainties in the frequency parameters on the generalized coherent states. The temporal degree of freedom in this mixed state is represented in the form of the power spectral density of the frequency comb. Once we have the general formalism, we compute the power spectral density of the frequency comb, by assuming that the carrier-envelop offset frequency and the pulse repetition frequency are random variables, having specific statistical uncertainties associated with them.

\section{Quantum formalism}
\label{kwantoes}

\subsection{Single-photon mixed state}

A mixed state is a convex sum over pure states
\begin{equation}
\hat{\rho} = \sum_n \ket{\psi_n}P_n\bra{\psi_n} ,
\end{equation}
where $\sum_n P_n=1$, so that $P_n$ is interpreted as probabilities. In a more general formalism, one can replace the summation with an integral
\begin{equation}
\hat{\rho} = \int \ket{\psi(\lambda)}P_{\lambda}(\lambda)\bra{\psi(\lambda)}\ {\rm d}\lambda ,
\label{mixdef}
\end{equation}
in which $\lambda$ is a continuous random variable that parameterizes the elements of the ensemble and $P_{\lambda}(\lambda)$ is a probability density function, such that
\begin{equation}
\int P_{\lambda}(\lambda)\ {\rm d}\lambda = 1 .
\label{pkon}
\end{equation}

To incorporate the frequency degree of freedom, we define the pure single-photon states by
\begin{eqnarray}
\ket{\psi(\lambda)} = \ket{1,G(\lambda)} & = & \int \ket{\nu} G(\nu;\lambda)\ {\rm d}\nu \nonumber \\
\bra{\psi(\lambda)} = \bra{1,G(\lambda)} & = & \int G^*(\nu;\lambda) \bra{\nu}\ {\rm d}\nu ,
\label{intoes}
\end{eqnarray}
where $G(\nu;\lambda)$ is the (stochastic) frequency spectrum or the Fourier domain wave function of the state. It is a function of the frequency $\nu$, which is related to the angular frequency by $\omega=2\pi\nu$. It is also a function of a random variable $\lambda$, which `labels' the elements of the ensemble. The bra- and ket-vectors $\bra{\nu}$ and $\ket{\nu}$ represent a one-dimensional frequency basis, which obeys the orthogonality condition $\braket{\nu_1}{\nu_2} = \delta(\nu_1-\nu_2)$. Using these definitions, we obtain the density operator for a single-photon
\begin{eqnarray}
\hat{\rho}_1 & = & \int \ket{1,G(\lambda)} P_{\lambda}(\lambda) \bra{1,G(\lambda)}\ {\rm d}\lambda \nonumber \\
& = & \int\!\!\!\int \ket{\nu} \left\langle G(\nu) G^*(\nu') \right\rangle \bra{\nu'}\ {\rm d}\nu\ {\rm d}\nu' ,
\label{eenmeng}
\end{eqnarray}
where $\langle \cdot \rangle$ denotes the ensemble average, so that
\begin{equation}
\left\langle G(\nu) G^*(\nu') \right\rangle = \int G(\nu;\lambda) G^*(\nu';\lambda) P_{\lambda}(\lambda)\ {\rm d}\lambda
\label{ensdef}
\end{equation}
is interpreted as the two-point correlation function in Fourier space. Although $\lambda$ is shown here as a single random variable, the expressions can be generalized to an arbitrary number of random variables.

One can convert the density operator into a density `matrix' in the Fourier basis by operating on both sides with the frequency basis elements
\begin{equation}
\rho_1(\nu,\nu') = \bra{\nu}\hat{\rho}_1\ket{\nu'} = \left\langle G(\nu) G^*(\nu') \right\rangle .
\label{digtpunt}
\end{equation}
The result indicates that the density matrix for the mixed single-photon state in the Fourier basis is precisely the two-point correlation function in the Fourier domain, obtained from the ensemble average.

The trace of the density operator
\begin{eqnarray}
{\rm tr}\{\hat{\rho}_1\} & = & \int \left\langle |G(\nu)|^2 \right\rangle\ {\rm d}\nu \nonumber \\
 & = & \int\!\!\!\int |G(\nu;\lambda)|^2 P_{\lambda}(\lambda)\ {\rm d}\lambda\ {\rm d}\nu = 1 ,
\label{eenmeng0}
\end{eqnarray}
represents the normalization condition for the stochastic spectra $G(\nu;\lambda)$. If we assume that
\begin{equation}
\int \left| G(\nu;\lambda) \right|^2 {\rm d} \nu = 1 ,
\label{glamnorm}
\end{equation}
for all elements of the ensemble (all values of $\lambda$), then it leads to Eq.~(\ref{pkon}), which then satisfies the normalization requirement.

\subsection{Stationary random process}

The spectra $G(\nu;\lambda)$ are related to real-valued stochastic time signals $g(t;\lambda)$ via the Fourier transform
\begin{equation}
G(\nu;\lambda) = \int g(t;\lambda) \exp(-\im 2\pi \nu t)\ {\rm d}t .
\label{ft}
\end{equation}
Usually, the time signals are treated as if they are of infinite duration. To avoid the divergences that such infinite duration signals can cause, we follow the standard approach of using a limit process, given by
\begin{equation}
\int_{-\infty}^{\infty} ...\ {\rm d}t \longrightarrow \lim_{T\rightarrow\infty} \frac{1}{T} \int_{-T/2}^{T/2} ...\ {\rm d}t \equiv \int_T ...\ {\rm d}t ,
\label{tlim}
\end{equation}
where we introduce a simplified notation to denote this limit process.

It turns out that, if the system is stationary, in that its statistical characteristics do not change with time, then the quantum states that are associated with time signals of infinite duration (and the two-point functions on which their definition is based) can be calculated in a well-defined manner, in terms of the limit process. The two-point function now becomes
\begin{eqnarray}
\left\langle G(\nu) G^*(\nu') \right\rangle & = & \int \int_T g(t_1;\lambda) \exp(-\im 2\pi \nu t_1)\ {\rm d}t_1 \nonumber \\
& & \times \int_T g(t_2;\lambda) \exp(\im 2\pi \nu' t_2)\ {\rm d}t_2\ P_{\lambda}(\lambda)\ {\rm d}\lambda \nonumber \\
& = & \int_T\!\int_T \left\langle g(t_1) g(t_2) \right\rangle \nonumber \\
& & \times \exp[-\im 2\pi (\nu t_1-\nu' t_2)]\ {\rm d}t_1\ {\rm d}t_2 ,
\end{eqnarray}
where
\begin{equation}
\left\langle g(t_1) g(t_2) \right\rangle = \int g(t_1;\lambda) g(t_2;\lambda) P_{\lambda}(\lambda)\ {\rm d}\lambda ,
\end{equation}
is the autocorrelation function.

If the random process is stationary (shift invariant), the autocorrelation function only depends on the difference between the variables $\langle g(t_1) g(t_2)\rangle = R_g(t_1-t_2)$. Let's redefine one of the variables by $t_1=t_2+\tau$. Then the expression becomes separable
\begin{eqnarray}
\left\langle G(\nu) G^*(\nu') \right\rangle
& = & \int_T \exp[-\im 2\pi (\nu-\nu') t_2]\ {\rm d}t_2 \nonumber \\
& & \times \int_T R_g(\tau) \exp[-\im 2\pi \nu \tau]\ {\rm d}\tau \nonumber \\
& = & \epsilon(\nu-\nu') S(\nu) ,
\label{korfdef}
\end{eqnarray}
where
\begin{eqnarray}
S(\nu)& = &\langle|G(\nu)|^2\rangle \nonumber \\
& = & \lim_{T\rightarrow\infty} \frac{1}{T} \int_{-T/2}^{T/2} R_g(\tau) \exp[-\im 2\pi \nu \tau]\ {\rm d}\tau
\label{psddef}
\end{eqnarray}
is the power spectral density, which (thanks to the Wiener-Khintchine theorem) is given by the Fourier transform of the autocorrelation function, and
\begin{equation}
\epsilon(\nu-\nu') \equiv \lim_{T\rightarrow\infty} \frac{1}{T} \int_{-T/2}^{T/2} \exp[\im 2\pi (\nu-\nu')t]\ {\rm d}t ,
\label{epsdef}
\end{equation}
is a special function, as defined through a limit process. In the limit, the $\epsilon$-function becomes
\begin{equation}
\epsilon(\nu) = \left\{ \begin{array}{ccc} 1 & {\rm for} & \nu=0 \\ 0 & {\rm for} & \nu \neq 0 \end{array} \right. .
\label{epsfunc}
\end{equation}
It is a function of measure zero, which implies
\begin{equation}
\int f(\nu) \epsilon(\nu-\nu')\ {\rm d}\nu = 0 ,
\end{equation}
for any function $f(\nu)$ that has a finite function value $f(\nu')$. Using the expression in Eq.~(\ref{epsfunc}), one can readily show that $\epsilon(-\nu)=\epsilon(\nu)$, $\epsilon^2(\nu)=\epsilon(\nu)$ and $\epsilon(a\nu)=\epsilon(\nu)$, where $a$ is an arbitrary constant.

It then follows that the density matrix is diagonal in the Fourier basis
\begin{equation}
\rho_1(\nu,\nu') = \epsilon(\nu-\nu') \langle|G(\nu)|^2\rangle = \epsilon(\nu-\nu') S(\nu) .
\label{eenmeng1}
\end{equation}
The lack of off-diagonal elements indicates that, as expected, there is no mutual coherence between different frequency components. The associated single-photon density operator, given by
\begin{equation}
\hat{\rho}_1 = \int\!\!\!\int \ket{\nu} \epsilon(\nu-\nu') S(\nu) \bra{\nu'}\ {\rm d}\nu\ {\rm d}\nu' ,
\label{eenmeng2}
\end{equation}
has a well-defined trace ${\rm tr}\{\hat{\rho}_1\}=1$, provided that
\begin{equation}
\int S(\nu) {\rm d} \nu = 1 ,
\end{equation}
which is implied by Eq.~(\ref{eenmeng0}).

The single-photon quantum state of the frequency comb can now be obtained by substituting the power spectral density into Eq.~(\ref{eenmeng2}). However, before we consider the power spectral density of the frequency comb, we develop the formalism further to address the multi-photon case.

\section{Multi-photon states}

\subsection{Mixed $n$-photon states}

A Fock state can be expressed in terms of a single-photon state that is raised to a given integer power. Using the definition of the pure single-photon state given in Eq.~(\ref{intoes}), we express the fixed-spectrum Fock state by
\begin{equation}
\ket{n,G} = \frac{1}{\sqrt{n!}} \left( \ket{1,G} \right)^n .
\label{genfock}
\end{equation}
Note that all the photons in this state share the same spectrum. The combinatoric factor is required for normalization. Generalizing it to mixed states, we proceed as before by assuming that the spectrum also depends on a random variable $G(\nu;\lambda)$. The mixed $n$-photon state is then defined in an analogous way as before, by
\begin{equation}
\hat{\rho}_n = \int \ket{n,G(\lambda)}P_{\lambda}(\lambda)\bra{n,G(\lambda)}\ {\rm d}\lambda .
\label{mixdefn}
\end{equation}

In terms of the definitions of the single-photon state in Eq.~(\ref{intoes}) and the Fock state in Eq.~(\ref{genfock}), the expression for the mixed $n$-photon state becomes
\begin{eqnarray}
\hat{\rho}_n & = & \frac{1}{{n!}} \int\! ...\!\int \ket{\nu_1}...\ket{\nu_n} \left\langle G(\nu_1)...G(\nu_n) G^*(\nu_1')...G^*(\nu_n') \right\rangle \nonumber \\
& & \times \bra{\nu_1'}...\bra{\nu_n'}\ {\rm d}\nu_1..{\rm d}\nu_n\ {\rm d}\nu_1'..{\rm d}\nu_n' ,
\label{mixdefn0}
\end{eqnarray}
where
\begin{eqnarray}
& & \left\langle G(\nu_1)...G(\nu_n) G^*(\nu_1')...G^*(\nu_n') \right\rangle) \nonumber \\ & = & \int G(\nu_1;\lambda)...G(\nu_n;\lambda) G^*(\nu_1';\lambda)...G^*(\nu_n';\lambda) P_{\lambda}(\lambda)\ {\rm d}\lambda , \nonumber \\
\label{npuntdef}
\end{eqnarray}
is a higher order correlation function.

At this point, we treat the function value of the spectrum at a particular frequency $G(\nu)$ as the random variable. If we assume that these random variables are normally distributed, then the ensemble average in Eq.~(\ref{npuntdef}) breaks up into a product of two-point functions. The only nonzero two-point functions are those that contain $G$ with its complex conjugate. There are $n!$ different ways to combine the $G$s with their complex conjugates. This gives a combinatoric factor of $n!$, which cancels the factor of $1/n!$ in Eq.~(\ref{mixdefn0}). All the two-point functions are equal and given by Eq.~(\ref{korfdef}). In the end, the expression becomes
\begin{eqnarray}
\hat{\rho}_n & = & \left[\int\!\!\!\int \ket{\nu} \epsilon(\nu-\nu') S(\nu) \bra{\nu'}\ {\rm d}\nu\ {\rm d}\nu'\right]^{\otimes n} \nonumber \\
& = & \left( \hat{\rho}_1 \right)^{\otimes n} .
\label{rhondef}
\end{eqnarray}
Under these circumstances, the density operator for the fixed-spectrum $n$-photon mixed state is given by the tensor product of $n$ single-photon mixed state density operators with the same spectrum.

\subsection{Bosonic nature of mixed $n$-photon states}

One may well ask, why should the $\lambda$-parameter for the different single-photon factors in an $n$-photon state all be the same, as we tacitly assumed in Eq.~(\ref{mixdefn})? The pure $n$-photon state could be expressed more generally as
\begin{equation}
\ket{n,G,\bar{\lambda}} = \frac{1}{\sqrt{n!}} \prod_{p=1}^n \ket{1,G(\lambda_p)} ,
\end{equation}
where $\bar{\lambda}=\{\lambda_p\}$ for $p=1...n$. The resulting density operator would then be given by
\begin{equation}
\hat{\rho}_n = \int \ket{n,G,\bar{\lambda}}P_{\bar{\lambda}}(\bar{\lambda})\bra{n,G,\bar{\lambda}}\ {\rm d}^n\lambda ,
\end{equation}
where $P_{\bar{\lambda}}(\bar{\lambda})$ is a $n$-dimensional joint probability density function, such that
\begin{equation}
\int P_{\bar{\lambda}}(\bar{\lambda})\ {\rm d}^n\lambda = 1 .
\end{equation}

Here, one can consider different scenarios. In the first scenario, the different $\lambda_p$'s are all statistically independent. In this case, the $n$-dimensional joint probability density function becomes the product of $n$ separate one-dimensional probability density functions. As a result, the integral breaks up into $n$ separate integrals, each representing a two-point correlation function. However, since each spectrum is only associated with one particular complex conjugate spectrum in this case, the combinatorics will not produce a factor of $n!$ to cancel the factor of $1/n!$. By implication, the result would be suppressed by a factor of $1/n!$.

An alternative scenario, which represents the extreme opposite, is where the different $\lambda_p$'s are perfectly correlated. In this case, the joint probability density function would be zero unless all the $\lambda_p$'s have the same value. This can be represented as a one-dimensional probability density function, multiplied by Dirac delta functions that set all the $\lambda_p$'s equal. The result is precisely the case that we considered in the previous section with the $2n$-point function depicted by Eq.~(\ref{npuntdef}). This scenario is more natural, because photons tend to exist in the same state due to their bosonic nature. This is supported by the enhancement that this scenario receives due to the cancellation of the $1/n!$-factor.

One can also have other scenarios that lies somewhere between these two extremes with enhancement factors ranging between those of the former and the latter scenarios. Since the latter scenario has the highest enhancement factor, we'll consider only this scenario.

\subsection{Mixed coherent states}

A coherent state can be generalized to include frequency degree of freedom by expressing it in terms of the fixed-spectrum Fock states of Eq.~(\ref{genfock}), leading to
\begin{equation}
\ket{\alpha,G} = \exp \left(-|\alpha|^2/2\right) \sum_{n=0}^{\infty} \frac{\alpha^n}{\sqrt{n!}}\ \ket{n,G} .
\label{fskoh}
\end{equation}
For the mixed case, we include the random variable $G(\nu)\rightarrow G(\nu;\lambda)$. Its density operator then reads
\begin{eqnarray}
\hat{\rho}_{\alpha} & = & \int \ket{\alpha,G(\lambda)}P_{\lambda}(\lambda)\bra{\alpha,G(\lambda)}\ {\rm d}\lambda \nonumber \\
& = & \exp \left(-|\alpha|^2\right) \int \sum_{m,n=0}^{\infty} \frac{\alpha^m(\alpha^*)^n}{\sqrt{m!}\sqrt{n!}} \nonumber \\
& & \times \ket{m,G(\lambda)} P_{\lambda}(\lambda)\bra{n,G(\lambda)}\ {\rm d}\lambda ,
\end{eqnarray}
when expressed in terms of Eq.~(\ref{fskoh}). Since the only nonzero two-point functions are those that contain $G$ with its complex conjugate, we have
\begin{equation}
\hat{\rho}_{mn} \equiv \int \ket{m,G(\lambda)} P_{\lambda}(\lambda)\bra{n,G(\lambda)}\ {\rm d}\lambda = \hat{\rho}_n \delta_{mn} ,
\end{equation}
where $\hat{\rho}_n$ is given in Eq.~(\ref{rhondef}). Hence, only the diagonal terms survive. The expression for the complete state becomes
\begin{eqnarray}
\hat{\rho}_{\alpha} & = & \exp \left(-|\alpha|^2\right) \sum_{m,n=0}^{\infty} \frac{\alpha^m(\alpha^*)^n}{\sqrt{m!}\sqrt{n!}}\ \hat{\rho}_{mn} \nonumber \\
& = & \exp \left(-|\alpha|^2\right) \sum_{n=0}^{\infty} \frac{|\alpha|^{2n}}{n!}\ \hat{\rho}_n .
\label{mixdefa}
\end{eqnarray}

In the end, one can write the density operator as an exponentiated operator. When we substitute Eq.~(\ref{rhondef}) into Eq.~(\ref{mixdefa}), we obtain
\begin{eqnarray}
\hat{\rho}_{\alpha} & = & \exp \left(-|\alpha|^2\right) \sum_{n=0}^{\infty} \frac{|\alpha|^{2n}}{n!} \nonumber \\
& & \times \left[ \int\!\!\!\int \ket{\nu} \epsilon(\nu-\nu') S(\nu) \bra{\nu'}\ {\rm d}\nu\ {\rm d}\nu' \right]^{\otimes n} \nonumber \\
& = & \exp \left(-|\alpha|^2\right) \exp_{\otimes} \left( |\alpha|^2 \hat{\rho}_1 \right) ,
\label{kankwa}
\end{eqnarray}
where $\exp_{\otimes}$ is defined such that all the operator products in its expansion are tensor products.

The quantum state of the frequency comb can now be expressed by the density operator in Eq.~(\ref{kankwa}), after substituting the expression of the power spectral density of the frequency comb into it. Next, we compute the power spectral density of the frequency comb.

\section{Power spectral density}

\subsection{Kerr-lens mode-locking}
\label{kerr}

The calculation of the power spectral density of a frequency comb, is based on the mechanism that generates a frequency comb. For this purpose, we consider the Kerr-lens mode-locking process \cite{kerrlock}.

The mechanism for Kerr-lens mode-locking involves a laser cavity that is designed such that loss in the cavity is minimized for high-intensity pulses that produce a Kerr-lensing effect. The different cavity modes add in-phase at a particular point in the cavity, which means that the difference in frequency between adjacent cavity modes (the mode separation) is constant over the spectral bandwidth of the laser. The latter places some requirements on the dispersion in the cavity, which is determined by the wavenumber as a function of frequency $k(\nu)$. Using a Taylor series expansion of the wavenumber about the carrier frequency $\nu_c$, one can distinguish among the different types of contributions, respectively associated with the phase velocity, the group velocity, the group velocity dispersion, and so forth. To have a constant mode separation, the group velocity dispersion and all higher order terms need to be zero. For Kerr-lens mode-locking, a special subsystem (using, for example, a pair of prisms \cite{dispcomp0}) is used to compensate for the group velocity dispersion. The effect of the remaining undesired terms could be reduced due to the effect of injection mode locking. Here we'll simply assume that all these undesired terms are zero. The phase velocity and group velocity then determine the mode spacing, which is given by the pulse repetition frequency $\nu_{\rm rep}$. They are also in part responsible for an offset (the carrier-envelop offset-frequency $\nu_{\rm ceo}$) between the lowest harmonic and zero that is not an integer multiple of the mode spacing given by $\nu_{\rm rep}$.

\subsection{Spectrum of a frequency comb}

The spectrum that is produced by the Kerr-lens mode locking mechanism can be expressed by
\begin{equation}
\tilde{\cal E}(\nu) = P(\nu-\nu_c) \sum_{m=-\infty}^{\infty} \delta(\nu-m \nu_{\rm rep}-\nu_{\rm ceo}) ,
\end{equation}
where $P(\nu)$ is an envelope function, representing the shape of the overall spectrum centered around the carrier frequency $\nu_c$ in both halves of the spectrum. The carrier-envelop offset-frequency $\nu_{\rm ceo}$, which represents the offset between the comb frequencies and the harmonic grid frequencies, is defined such that $0<\nu_{\rm ceo}<\nu_{\rm rep}$.

Note that $\tilde{\cal E}(\nu)$ is a one-sided spectrum. The full spectrum is defined by
\begin{equation}
\tilde{E}(\nu) = \frac{1}{2} [\tilde{\cal E}(\nu)+\tilde{\cal E}^*(-\nu)] .
\label{volspek}
\end{equation}
One can allow the summations of the spectral components in the respective one-sided spectra to run from $-\infty$ to $\infty$, because the additional Dirac delta functions will fall outside the region where $P(\nu \pm \nu_c)$ is nonzero and thus won't contribute. Since the time-signal associated with the full spectrum is a real valued function, we have that $P^*(\nu)=P(-\nu)$.

\subsection{Ensemble averaging}
\label{ensav}

The power spectral density is the modulus square of the spectrum. For this purpose, we need to add both sides of the spectrum as in Eq.~(\ref{volspek}). However, to compute the power spectral density, we need to treat the Dirac delta functions with care. In practice, the pulse train would be of finite duration. As such it is multiplied by an overall envelop function that limits the time duration of the pulse train on the time domain. The comb spectrum is convolved with a narrow function $h(\nu)$, converting the Dirac delta functions into narrow spectral component functions. The envelop function and its spectrum $h(\nu)$ are finite energy functions. After taking the modulus square of the spectrum, one can make the calculation more manageable by converting the squares of the narrow functions back into Dirac delta functions $h(\nu)^2\rightarrow\Delta\delta(\nu)$, with the understanding that one would in the process pick up a factor of an extra dimension parameter $\Delta$. This dimension parameter can be absorbed into $P(\nu)$ so that we don't need to show it explicitly. The result can be expressed as
\begin{eqnarray}
S(\nu) & = & \frac{1}{4} |P(\nu-\nu_c)|^2 \sum_{m=-\infty}^{\infty} \delta(\nu - \nu_{\rm ceo} - m \nu_{\rm rep}) \nonumber \\
& & + \frac{1}{4}  |P(\nu+\nu_c)|^2 \sum_{m=-\infty}^{\infty} \delta(\nu + \nu_{\rm ceo} + m \nu_{\rm rep}) \nonumber \\
& = & \frac{1}{4} \left[ {\cal S}(\nu) + {\cal S}(-\nu) \right] .
\label{volspek0}
\end{eqnarray}
Note that the cross terms fall away, because they don't overlap on the frequency domain.

Information about the coherence of a laser source is contained in its power spectral density; the inverse Fourier transform of the power spectral density is the mutual coherence function. The coherence of the frequency comb is affected by the statistical properties of the carrier-envelop offset frequency $\nu_{\rm ceo}$ and the pulse-repetition frequency $\nu_{\rm rep}$. To take the statistical properties of these quantities into account, we treat them as random variables and evaluate the power spectral density as an ensemble average. Here, this is done for the positive frequency term
\begin{eqnarray}
{\cal S}(\nu) & = & \langle |\tilde{\cal E}(\nu)|^2 \rangle \nonumber \\
& = & \left\langle |P(\nu-\nu_c)|^2 \sum_{m=-\infty}^{\infty} \delta(\nu - \nu_{\rm ceo} - m \nu_{\rm rep}) \right\rangle . \nonumber \\
\end{eqnarray}
Expressing the Dirac delta function in terms of its Fourier transform, we obtain
\begin{equation}
{\cal S}(\nu) = |P(\nu-\nu_c)|^2 \sum_{m=-\infty}^{\infty} Q_m(\nu) ,
\label{halfpsd}
\end{equation}
where
\begin{equation}
Q_m(\nu) = \int \exp(\im 2\pi \xi \nu) \langle \exp[-\im 2\pi \xi(\nu_{\rm ceo} + m \nu_{\rm rep})] \rangle\ {\rm d}\xi .
\end{equation}
Here, $\xi$ is an auxiliary integration variable. Since the random variables only appear in the exponential function under the $\xi$-integral, one can restrict the ensemble averaging to this exponential function. By assuming that the two random variables are statistically independent, one can separate the ensemble average into the product of two ensemble averages and evaluate them separately. We also assume that the random variables are normally distributed, so that one can express their probability density functions by
\begin{equation}
p_X(\nu_X) = \frac{1}{\sqrt{2\pi}\sigma_X} \exp\left[\frac{-(\nu_X - \mu_X)^2}{2\sigma_X^2}\right] ,
\end{equation}
where $\mu_X$ is the mean of the distribution and $\sigma_X$ is its standard deviation. The subscript $X$ can either denote `ceo' or `rep' to represent the carrier-envelop offset frequency or the pulse-repetition frequency, respectively.

After computing the ensemble average, we find
\begin{eqnarray}
& & \langle \exp[-\im 2\pi \xi(\nu_{\rm ceo} + m \nu_{\rm rep})] \rangle \nonumber \\ & = & \exp[-\im 2\pi \xi(\mu_{\rm ceo}+ m \mu_{\rm rep})] \nonumber \\
& & \times \exp[-2\pi^2 \xi^2(\sigma_{\rm ceo}^2 + m^2 \sigma_{\rm rep}^2)] .
\end{eqnarray}
The integration over $\xi$ then leads to
\begin{eqnarray}
Q_m(\nu) & = & \frac{1}{\sqrt{2\pi (\sigma_{\rm ceo}^2 + m^2 \sigma_{\rm rep}^2)}} \nonumber \\
& & \times \exp \left[ \frac{ -(\nu - \mu_{\rm ceo} - m \mu_{\rm rep})^2}{2 (\sigma_{\rm ceo}^2 + m^2 \sigma_{\rm rep}^2)} \right] ,
\end{eqnarray}
which we can substitute into the power spectral density for the positive side, given in Eq.~(\ref{halfpsd}). A similar calculation produces the negative side of the spectrum. The full power spectral density, according to Eq.~(\ref{volspek0}), is given by
\begin{eqnarray}
S(\nu) & = & \frac{1}{4} \sum_{m=-\infty}^{\infty} \frac{|P(\nu-\nu_c)|^2}{\sqrt{2\pi (\sigma_{\rm ceo}^2 + m^2 \sigma_{\rm rep}^2)}} \nonumber \\
& & \times \exp \left[ \frac{ -(\nu - \mu_{\rm ceo} - m \mu_{\rm rep})^2}{2 (\sigma_{\rm ceo}^2 + m^2 \sigma_{\rm rep}^2)} \right] \nonumber \\
& & + \frac{1}{4} \sum_{m=-\infty}^{\infty} \frac{|P(\nu+\nu_c)|^2}{\sqrt{2\pi (\sigma_{\rm ceo}^2 + m^2 \sigma_{\rm rep}^2)}} \nonumber \\
& & \times \exp \left[ \frac{ -(\nu+\mu_{\rm ceo}+m\mu_{\rm rep})^2}{2 (\sigma_{\rm ceo}^2 + m^2 \sigma_{\rm rep}^2)} \right] .
\label{kampsd}
\end{eqnarray}
The result consists of Gaussian components that become progressively broader as $m$ increases.

Substituting Eq.~(\ref{kampsd}) into Eq.~(\ref{eenmeng2}), we obtain an expression for the single-photon quantum state of a frequency comb. Upon substituting the result into Eq.~(\ref{kankwa}), one obtains the expression for the multi-photon quantum state of a frequency comb.

\subsection{Example calculation}
\label{examp}

\begin{figure}[th]
\includegraphics{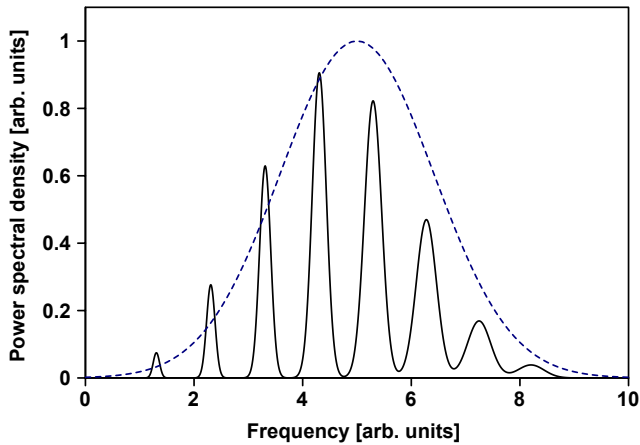}
\caption{Shape of the power spectral density function (black line) is plotted for parameters $\mu_{\rm ceo}=0.3$, $\mu_{\rm rep}=1$, $\sigma_{\rm ceo}=0.05$ and $\sigma_{\rm rep}=0.03$ in arbitrary units. For comparison, the Gaussian envelop function is shown (blue dashed line) with $\nu_c=5$ and $B=2$ in arbitrary units.}
\label{psd}
\end{figure}

To see what the power spectral density looks like, we provide a curve in Fig.~\ref{psd}, where we selected parameter values that, although not realistic, would demonstrate their effect on the curve. For this purpose, we model the envelop function as a Gaussian function
\begin{equation}
P(\nu-\nu_c) = \exp\left[ -\frac{(\nu-\nu_c)^2}{B^2}\right] ,
\label{modelenv}
\end{equation}
where $\nu_c=5$ and $B=2$ in arbitrary units. The remaining parameters are chosen as $\mu_{\rm ceo}=0.3$, $\mu_{\rm rep}=1$, $\sigma_{\rm ceo}=0.05$ and $\sigma_{\rm rep}=0.03$ in the same arbitrary units.

One can identify the individual frequency components in the frequency comb, broadened by the uncertainties in $\nu_{\rm ceo}$ and $\nu_{\rm rep}$. The uncertainty in $\nu_{\rm rep}$ causes the broadening to increase for higher frequency components. At the same time, these components are suppressed relative to the components at lower frequencies, as can be seen by comparing the peak amplitudes of these components to the shape of the envelop function, shown as the blue dashed curve in Fig.~\ref{psd}.

\section{Discussion}
\label{disc}

The approach that is used to develop the required quantum formalism involves the frequency basis vector states. One could alternatively use creation and annihilation operators to formulate equivalent expressions for the required formalism. Our choice to express it in terms of basis vectors, is purely for the sake of convenience.

When we introduce the limit process to deal with time signals of infinite duration, we always consider the result as if we have applied the limit. In practice, the signals are never actually of infinite duration; measurements are made with finite integration times. The correspondence between the theoretical limit and the practical finite integration time comes from an assumption that the limiting value would be reached after a long enough finite integration period. However, this is often not a valid assumption. For instance, the statistical properties of the different kinds of noise that affects the fractional frequency uncertainty (or instability) of laser sources follow particular power laws, which lead to a nontrivial dependence on the integration time. As a result, the fractional frequency uncertainty is often represented by the Allan deviation \cite{allan}, plotted as a function of the logarithmic integration time to reveal how it behaves over different scales of the integration period. The formulation that we provide here can be altered to accommodate this requirement. Instead of considering the power spectral density as the limiting case, one can forego taking the limit in Eq.~(\ref{psddef}) and retain the integration period as an additional parameter $S(\nu)\rightarrow S(\nu;T)$. Apart from this replacement, the formalism would remain the same; we still apply the limit in Eq.~(\ref{epsdef}) to obtain the $\epsilon$-function.

We want to emphasize the significance of the expression in Eq.~(\ref{eenmeng2}). The trace of an identity operator, defined over a continuous basis, is inevitably divergent. Therefore, an attempt to define a density operator for a completely mixed state in such a continuous basis, as a generalization of the identity operator would result in a density operator that cannot be normalized:
\begin{equation}
{\rm tr}\left\{\int \ket{\nu} S(\nu) \bra{\nu}\ {\rm d}\nu \right\} = \int \delta(0) S(\nu)\ {\rm d}\nu .
\end{equation}
The expression in Eq.~(\ref{eenmeng2}) circumvents this problem, thanks to the inclusion of the $\epsilon$-function. Moreover, since multi-particle states are build up by starting from the definition of the single-particle state, the expression in Eq.~(\ref{eenmeng2}) forms the foundation of the formalism for multi-particle states that are maximally mixed in terms of an extra degree of freedom.

The final expression for the mixed coherent state in Eq.~(\ref{kankwa}) is parameterized by the complex parameter $\alpha$, which is associated with the particle-number degrees of freedom, and the power spectral density $S(\nu)$, which is associated with the frequency degree of freedom. It is interesting to note that the way these quantities appear in the final expression allows one to combine them into a single complex function $\eta(\nu)=\alpha G(\nu)$. One can then, for instance, define an operator
\begin{equation}
|\alpha|^2 \hat{\rho}_1 = \hat{\sigma}_1 = \int\!\!\!\int \ket{\nu} \epsilon(\nu-\nu') \langle|\eta(\nu)|^2\rangle \bra{\nu'}\ {\rm d}\nu\ {\rm d}\nu' ,
\label{eenmeng3}
\end{equation}
so that its trace is given by ${\rm tr}\{\hat{\sigma}_1\} = |\alpha|^2$. The expression for the mixed coherent state then becomes $\hat{\rho}_{\alpha} = \exp \left( -{\rm tr}\{\hat{\sigma}_1\}\right) \exp_{\otimes} \left(\hat{\sigma}_1\right)$. As such the particle-number degrees of freedom and the frequency degree of freedom are combined and represented by one complex function.

The formalism that we obtain, as expressed in Eq.~(\ref{kankwa}), is specifically developed to express the quantum state of the frequency comb in terms of both the particle-number degrees of freedom and the frequency degrees if freedom, given the uncertainties in its frequency parameters. In this analysis, the particle-number degrees of freedom retain their characteristics as that of a coherent state; we do not consider any processes such as squeezing. It should therefore not be regarded as a general formalism to express mixed states in terms of multiple continuous degrees of freedom. The expression is expected to change if the particle-number degrees of freedom are manipulated through a process that alters its characteristics beyond that for a coherent state. Nevertheless, the current approach may allow one to develop the formalism further to accommodate such more general conditions.

\section{Conclusion}
\label{concl}

An expressed is derived for the quantum state of a frequency comb, in terms of both its particle-number degrees of freedom and its frequency degree of freedom, where the latter is parameterized by its power spectral density. We specifically considered the effect of uncertainties in the carrier-envelop offset frequency and the pulse repetition frequency to compute the power spectral density. These uncertainties give rise to mixing in the quantum state with the result that the frequency comb need to be expressed as a mixed state. To express such a mixed quantum state that depends on frequency as a degree of freedom, in addition to the particle-number degrees of freedom, we develop a specific quantum representation of the density matrix that incorporates both these degrees of freedom.

The resulting power spectral density shows the effect of uncertainties in the respective frequency parameters. This is demonstrated by an example.

\section*{Acknowledgements}

The research was done with the partial support of a grant from the National Research Foundation (NRF).


\end{document}